\documentclass[%
reprint,
groupedaddress,
 amsmath,amssymb,
 aps,
]{revtex4-1}

\usepackage{graphicx}
\usepackage{dcolumn}
\usepackage{bm}
\usepackage{hyperref}
\usepackage{siunitx}
\usepackage{comment}

\begin{document}


\title{Two-component dynamics and the liquid-like to gas-like crossover in supercritical water}


\author{Peihao Sun}
 \altaffiliation[Also at]{ Stanford University Physics Department, 382 Via Pueblo Mall, Stanford, CA 94305, USA}
 \email{phsun@stanford.edu}
\author{J. B. Hastings}%
\affiliation{%
 SLAC National Accelerator Laboratory, 2575 Sand Hill Rd, Menlo Park, CA 94025, USA
}%

\author{Daisuke Ishikawa}
\author{Alfred Q. R. Baron}
\affiliation{
 Materials Dynamics Laboratory, RIKEN SPring-8 Center, 1-1-1 Kouto, Sayo, Hyogo 679-5148, Japan
}%

\author{Giulio Monaco}
\affiliation{%
 Dipartimento di Fisica, Universit{\`a} di Trento, I-38123 Povo (Trento), Italy
}%

\date{\today}

\begin{abstract}

Molecular-scale dynamics in sub- to super-critical water is studied with inelastic X-ray scattering and molecular dynamics simulations. The obtained longitudinal current correlation spectra can be decomposed into two main components: a low-frequency (LF), gas-like component and a high-frequency (HF) component arising from the O--O stretching mode between hydrogen-bonded molecules, reminiscent of the longitudinal acoustic mode in ambient water. With increasing temperature, the hydrogen-bond network diminishes and the spectral weight shifts from HF to LF, leading to a transition from liquid-like to gas-like dynamics with rapid changes around the Widom line.


\end{abstract}

\maketitle


The supercritical state of water was discovered almost 200 years ago~\cite{CagniarddelaTour1822}, but there were few subsequent studies for a long time due to experimental challenges. In recent years, however, interest in supercritical water has been growing. This is not only because of its natural occurrence around hydrothermal vents~\cite{Simoneit1993, Martin2008} and in the Earth's mantle~\cite{Hirschmann2012} which are of biological and geological importance, respectively, but also because of its wide application in bio-chemical technologies such as green materials synthesis~\cite{Adschiri2011}, biofuel production~\cite{Peterson2008}, and industrial waste treatment~\cite{Bermejo2006}. Underlying many of these applications is the large tunability of density and solvation properties near the critical point~\cite{Eckert1996, Peterson2008}. For water in particular, as temperature increases in the near-critical region, its solubility for inorganic ionic compounds decreases rapidly, while many simple organic compounds and gases become soluble or even completely miscible in supercritical water~\cite{Peterson2008}. These properties have been and can be further exploited to facilitate a variety of chemical processes~\cite{Akiya2002, Peterson2008}.

From a physical viewpoint, supercritical fluids have attracted attention recently because evidence suggests the existence of ``liquid-like'' and ``gas-like'' states beyond the critical point~\cite{Cunsolo1998, Gorelli2006}. A number of recent studies have thus focused on the transition between the two; in particular, the ``Widom line'' (WL), defined as the line of maximum correlation length in the supercritical region~\cite{Xu2005}, has been proposed as a liquid/gas separatrix~\cite{McMillan2010, Simeoni2010, Gallo2014, Banuti2015}. Another school of thought focuses on dynamics and proposes a separate ``Frenkel line'' as the border between liquid-like and gas-like states~\cite{Trachenko2016, Yang2015}. Still another viewpoint, specific to water, focuses on the hydrogen-bond (H-bond) network structure and its related percolation threshold~\cite{Partay2005, Bernabei2008}. Meanwhile, the uniqueness of these boundaries has been called into question~\cite{Schienbein2018}. Therefore, in this study, we elucidate the liquid-like to gas-like crossover in supercritical water by investigating {\AA}ngstr{\"o}m-scale dynamics \emph{via} inelastic X-ray scattering (IXS) and molecular dynamics (MD) simulations.

IXS measurements were carried out at BL43LXU~\cite{Baron2010} of the RIKEN SPring-8 Center in Japan. We used the Si(999) reflection with \SI{17.8}{keV} incident X-rays. The resolution function was measured with a \SI{2}{\mm} plexiglass sample and had a full width at half maximum of approximately \SI{3}{\meV} for all of the 20 analyzers used in the experiment.
The sample length was \SI{1.85}{\mm} in a pressure cell with diamond windows specifically designed for use with supercritical water~\cite{Ishikawa2012-sp}. We scanned from \SI{-40}{\meV} to \SI{40}{\meV} in photon energy transfer, and each scan took approximately \SI{1.5}{hrs}, during which the temperature and pressure were controlled within $\pm \SI{5}{K}$ and $\pm \SI{10}{bar}$. Background from the empty cell with windows was measured and subtracted, though it was negligible in the frequency range of interest.

\begin{figure}
    \includegraphics[width=0.45\textwidth]{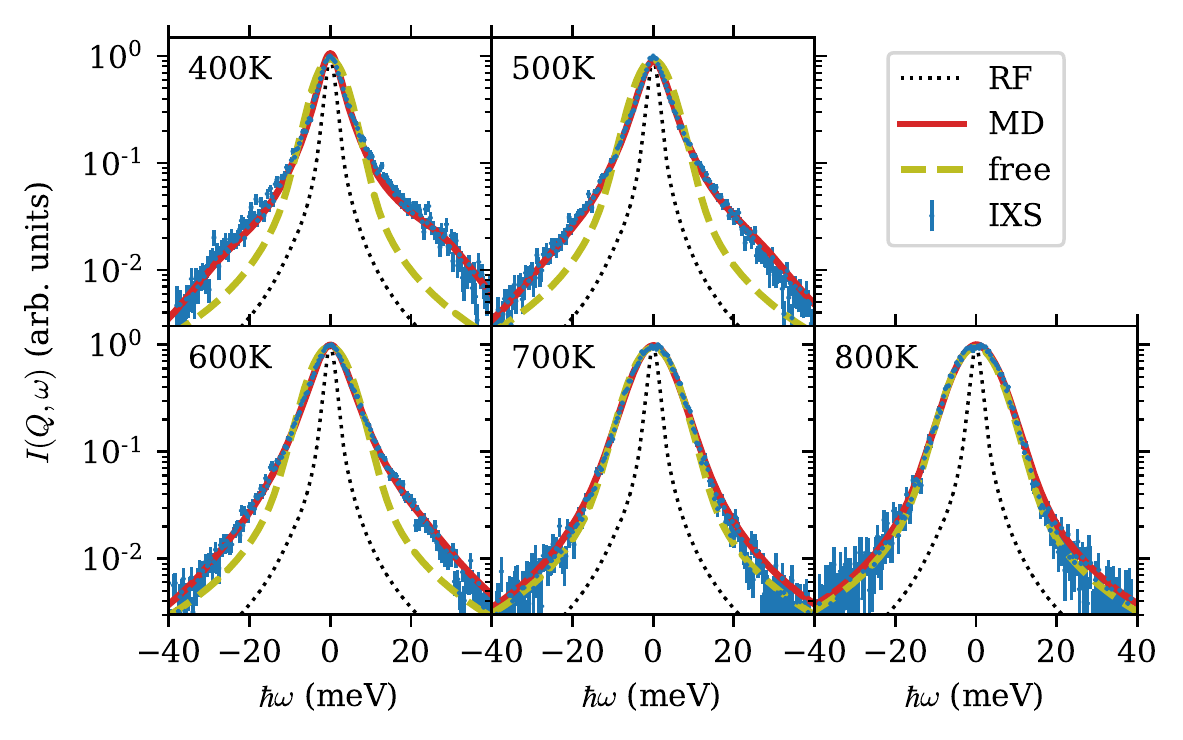}
    \caption{Comparison of IXS data (dots with error bar) and MD simulation results (solid line) at \SI{300}{bar} for $Q\approx \SI{12}{nm^{-1}}$. The MD results are multiplied by the Bose factor and convolved with the RF (dotted line). The free-gas limit (dashed line) is also plotted for comparison. All spectra are scaled to match at the center ($\omega=0$).}
    \label{fig:compare_MD_IXS}
\end{figure}

MD simulations are carried out using the LAMMPS simulation package~\cite{Plimpton1995}. We use $NPT$ ensembles with 2880 water molecules. After equilibration at each $P$-$T$ state, the simulation is run for \SI{1}{ns} at \SI{1}{fs} time steps. We choose the TIP4P/2005 potential~\cite{Abascal2005}, which is shown to best reproduce the physical properties of water among similar models~\cite{Vega2011}; in particular, the critical temperature and density
are closest to experimental values.
Even though its critical pressure, $P_c^\text{MD}=\SI{146}{bar}$, is lower than the experimental value $P_c=\SI{221}{bar}$ by \SI{75}{bar}, it is found that the TIP4P/2005 model can reproduce many dynamic and thermodynamic properties of supercritical water with a simple pressure change~\cite{Gallo2014}. Therefore, we apply a \SI{75}{bar} shift in our simulations to match experimental conditions, and this shift is implied in the discussions below.

In IXS, after background substraction, the scattered intensity at momentum transfer $\hbar Q$ and energy transfer $\hbar \omega$ is~\cite{Monaco1999, Baron2020-sp}:
\begin{equation} \label{eqn:IXS_int}
     I(Q, \omega) = I_0 [R(\omega)] * [B(\omega)S(Q, \omega)].
\end{equation}
Here $*$ denotes convolution; $I_0$ is an overall intensity factor; $R(\omega)$ is the instrument resolution function (RF); $B(\omega)$ is the Bose factor arising from detailed balance, which for liquids is generally taken to be $B(\omega)=(\hbar \omega /k_B T) / (1-e^{-\hbar \omega /k_B T})$~\cite{Monaco1999, Baron2020-sp}; $S(Q, \omega)$, the \emph{classical} dynamic structure factor, is the main quantity we are concerned with. Because the IXS signal is dominated by scattering from oxygen atoms, we include only oxygen atoms when calculating $S(Q, \omega)$ from the MD trajectories.

For ease of discussion, we first focus on the isobar $P=\SI{300}{bar}$ (\SI{225}{bar} for MD). Figure~\ref{fig:compare_MD_IXS} presents a comparison between the measured IXS spectra and MD simulation results calculated from Eq.~(\ref{eqn:IXS_int}). Features in the IXS data are well reproduced by the MD model. In particular, at \SI{400}{K}, both MD and IXS results show clear side bumps around $\pm\SI{25}{\meV}$. At $Q=\SI{12}{\per\nm}$, they correspond to a sound speed of approximately \SI{3}{km/s}, almost twice as fast as the ultrasonic value \SI{1.57}{km/s}. This is similar to what has been measured before in ambient water~\cite{Teixeira1985, Sette1995} and hence suggests liquid-like dynamics. With increasing temperature, these side bumps diminish while the central peak broadens. By \SI{800}{K}, the spectrum becomes very close to the ideal-gas limit (see e.g.\ Section~4.3 of Ref.~\onlinecite{Boon1991}):
\begin{equation} 
    S(Q, \omega) = \frac{1}{\sqrt{2\pi} Q v_0} \exp\left[ -\frac{1}{2}\left( \frac{\omega}{Q v_0}\right)^2 \right],
\end{equation}
where $v_0 \equiv \sqrt{k_B T/M}$ is the thermal velocity and $M$ the molecular mass of water.

\begin{figure}
    \includegraphics[width=0.48\textwidth]{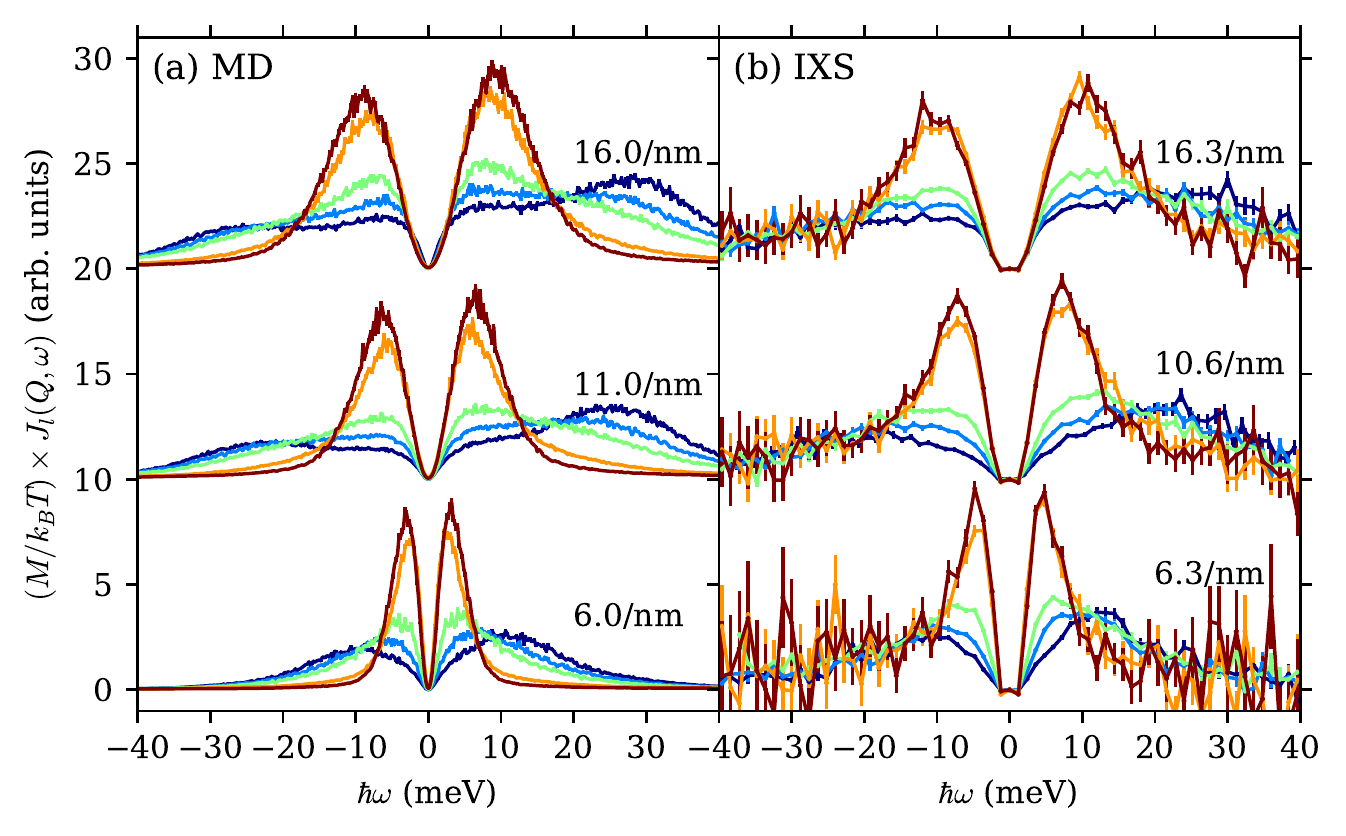}
    \caption{Longitudinal current correlation function, $J_l(Q, \omega)$, for $P=\SI{300}{bar}$ obtained from (a) MD and (b) IXS. MD spectra are multiplied by the Bose factor. IXS spectra are multiplied by $\omega^2/Q^2$ after subtracting out a scaled RF (see Supplemental Material). The spectra from dark blue to dark red are taken from \SI{400}{K} to \SI{800}{K} at \SI{100}{K} steps. An offset is applied between different spectra at different $Q$ values as indicated in the plots.}
    \label{fig:JL_300bar}
\end{figure}

To better investigate this liquid-like to gas-like transition in the dynamics, we focus on the longitudinal current-current correlation function $J_l(Q, \omega)$, which is often used to investigate acoustic modes at finite wavelengths. It can be shown that $J_l(Q, \omega)$ bears a simple relation to $S(Q, \omega)$~\cite{Boon1991}:
\begin{equation} \label{eqn:J_S_relation}
    J_l(Q, \omega) = \frac{\omega^2}{Q^2} S(Q, \omega).
\end{equation}
Moreover, because of the sum rule~\cite{Boon1991} $\int_{-\infty}^\infty J_l(Q, \omega) d\omega = k_B T / M$, $J_l(Q, \omega)$ can be conveniently normalized with a factor $M/k_B T$, making it easy to quantify spectral components as will be shown below.
The normalized $J_l$ spectra are presented in Fig.~\ref{fig:JL_300bar}. The MD spectra are multiplied by the Bose factor which reproduces the enhanced IXS signal on the Stokes ($\omega>0$) side. To obtain $J_l$ from IXS data, we multiply the IXS intensities by $\omega^2/Q^2$ and subtract the quasi-elastic background which becomes significant near the critical point (see Supplemental Material). Again, details of the temperature evolution in the IXS data are well reproduced in the MD results. 

Upon closer examination, two distinct components can be seen in $J_l$
: a low frequency (LF) component peaked below \SI{10}{\meV}, and a high frequency (HF) component peaked around \SI{25}{\meV}. While the HF component corresponds to the side bumps at lower temperatures in Fig.~\ref{fig:compare_MD_IXS}, thus representing liquid-like dynamics, the LF component eventually becomes the ideal-gas-like spectrum at high temperatures. With increasing temperature, the spectral weight shifts from HF to LF, leading to a crossover from liquid-like to gas-like dynamics.

Having established the phenomenology for the dynamic crossover, we now quantify this transition and find its physical origin. To our knowledge, however, no theory to date can fully reproduce the spectra shown in Fig.~\ref{fig:JL_300bar}. Fitting with given spectral shapes such as damped harmonic oscillators~\cite{Teixeira1985, Sette1995, Giordano2010} or the memory function~\cite{Monaco1999, Yamaguchi2005, Bencivenga2007a}, as is usually done in IXS data analyses, results in large fluctuations in the fit parameters.
In addition, the physical motivation behind these fit models---the existence of acoustic modes---is lost at high temperatures when the fluid becomes gas-like. Therefore, we take a phenomenological approach and use non-negative matrix factorization (NMF), which is model-independent and has been shown to give parts-based representations of data when properly constrained~\cite{Hoyer2004}. Mathematically, we optimize the decomposition
\begin{equation} \label{eqn:NMF_def}
  \begin{split}
        J_l(Q, \omega; P, T) = & f_\text{HF}(P, T) J_l^\text{HF}(Q, \omega)  \\
          & + f_\text{LF}(P, T) J_l^\text{LF}(Q, \omega) 
  \end{split}
\end{equation}
and find the $PT$-independent components $J_l^\text{HF}$ and $J_l^\text{LF}$, as well as their weights $f_\text{HF}$ and $f_\text{LF}$ for each thermodynamic state. The NNDSVD initialization~\cite{Boutsidis2008} is used to facilitate the separation of the components, although other initializations in Ref.~\onlinecite{Boutsidis2008} lead to essentially the same results.
Data below \SI{6}{\per\nm} are excluded to avoid the influence from critical fluctuations (large quasi-elastic scattering intensities as seen in, e.g., Ref.~\onlinecite{Yamaguchi2005} and~\onlinecite{Bencivenga2007a}). The data in the $Q$ range \SIrange{6}{18}{\per\nm} are fit simultaneously, i.e.\ with the same weights $f_\text{HF}$ and $f_\text{LF}$. This is found to give consistent results. Moreover, because of the aforementioned sum rule for $J_l$, we may normalize the components $J_l^\text{HF}$ and $J_l^\text{LF}$ so they both have the same area as $J_l$. Then, $f_\text{HF} + f_\text{LF} = 1$ for each $PT$ state, and in the following discussion we use $f \equiv f_\text{HF}$ to denote the fraction of the HF component. We note that this decomposition scheme is conceptually similar to the two-phase thermodynamic model proposed by Lin et al.~\cite{Lin2003, Lin2010} and the low-frequency fraction $f_\text{LF}=1-f$ is analogous to the translational ``fluidicity'' parameter defined therein. However, our method avoids the need to find a reference hard-spheres system and, as discussed below, the ability to examine the $Q$-dependence helps provide insight into the nature of the LF and HF components.

\begin{figure}
    \includegraphics[width=0.48\textwidth]{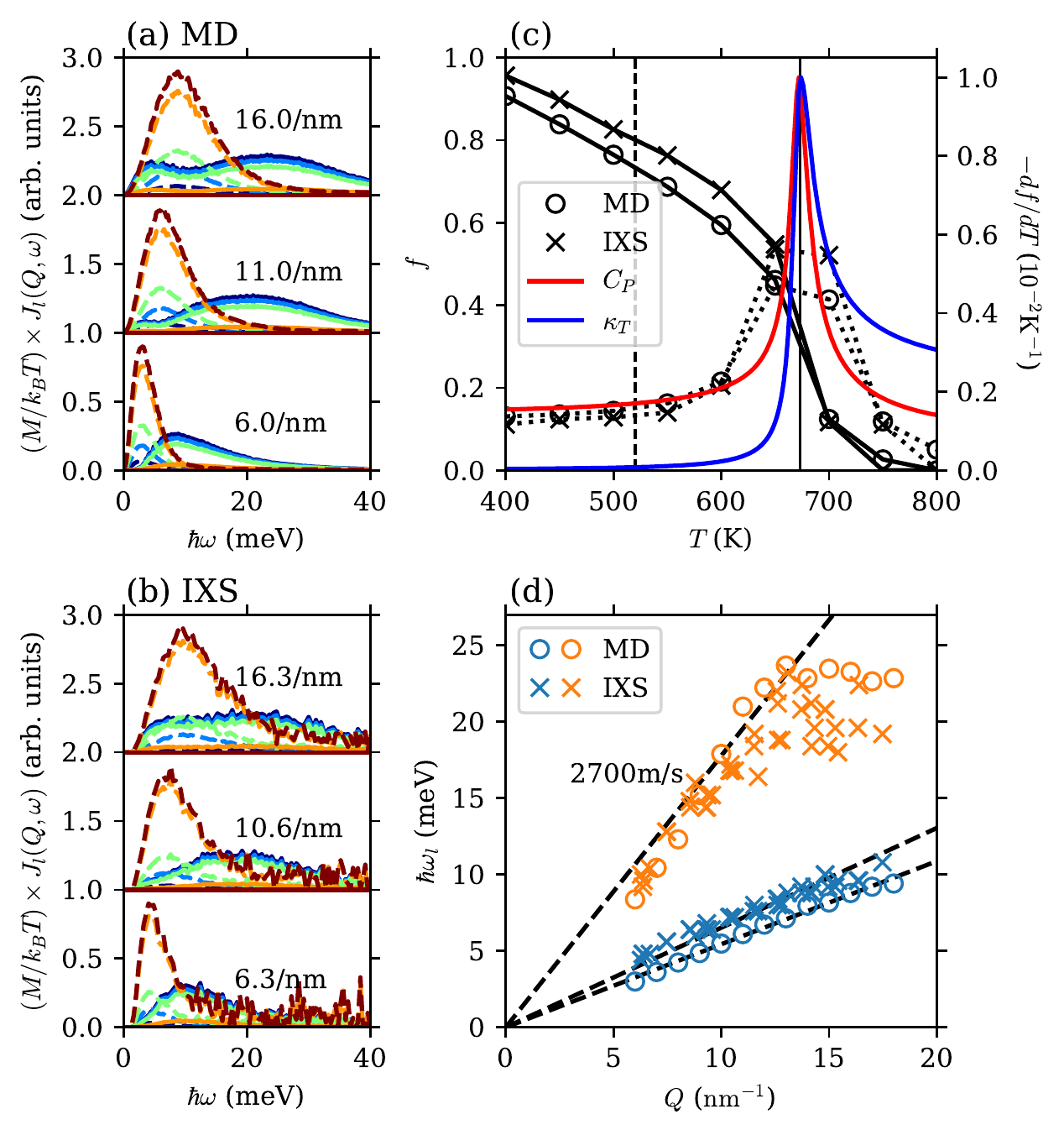}
    \caption{NMF results. (a) Dashed and solid lines indicate the LF and HF components of the MD spectra; (b) the same for IXS spectra. Color schemes and $Q$ values are the same as in Fig.~\ref{fig:JL_300bar}. (c) Circles: MD; crosses: IXS. Black solid lines show fraction of the HF component as a function of temperature (see text), while dotted lines show their temperature derivatives; these lines are guide to the eye. Also plotted are $C_P$ (red) and $\kappa_T$ (blue) scaled to match the derivatives. Solid and dashed vertical lines indicate the approximate position of the Widom line (maximum of $C_P$ and $\kappa_T$) and the Frenkel line~\cite{Yang2015}, respectively. (d) Dispersion curves for peak frequencies of the LF (blue) and HF (orange) components. Symbols are the same as in (c); dashed lines indicate linear dispersions.
    }
    \label{fig:f_vs_T}
\end{figure}

Figure~\ref{fig:f_vs_T}(a) and (b) present the NMF results for the spectra shown in Fig.~\ref{fig:JL_300bar} after dividing out the Bose factor. The shapes and temperature dependence of both components are discernible. In Fig.~\ref{fig:f_vs_T}(c), it can be seen that the fraction of the HF component $f$ obtained from MD and IXS closely follow each other, and both decrease monotonically as temperature increases. The presence of the WL is visible as a sudden drop in $f$
as temperature rises. This becomes clearer when we look at the derivative $|df/dT|$, calculated using the finite difference approximation, and compare it with thermodynamic response functions such as isobaric heat capacity $C_P$ and isothermal compressibility $\kappa_T$. Thus, $f$ can be used to describe the dynamic crossover in the supercritical region and is tied closely to thermodynamic properties. Because the analysis and discussion so far are not specific to water, the close connection between dynamics and thermodynamics may be applicable to other fluid systems exhibiting the two-component behavior. A possible candidate is liquid Te which, similar to water, shows a large positive sound dispersion close to its melting point~\cite{Kajihara2008}. We note in passing that the Frenkel line given in Ref.~\onlinecite{Yang2015} does not appear to correspond to significant changes in the dynamics, as shown in Fig.~\ref{fig:f_vs_T}(c).

In order to identify the microscopic origin of the dynamic transition, it is helpful to examine the dispersion relations $\omega_l(Q)$ for the HF and LF components, where the mode frequency $\omega_l$ is defined as the peak frequency at a given $Q$. The results are shown in Fig.~\ref{fig:f_vs_T}(d). For the LF component, $\omega_l$ changes linearly with $Q$, as expected for an ideal gas (see Supplemental Material). The HF component, on the other hand, has a $Q$-dependence characteristic of acoustic modes: it disperses linearly with $Q$ until it flattens around \SI{12}{\nm^{-1}}, i.e., near the boundary of the pseudo-Brillouin zone~\cite{Giordano2010}, which closely resembles the longitudinal acoustic branch observed in ambient liquid water~\cite{Sette1995, Sampoli1997}. Notice also that the peak frequency near the zone boundary is around \SI{25}{\meV} ($\SI{200}{\cm^{-1}}$), recalling earlier Raman and far-IR measurements in which a peak at similar position was seen~\cite{Walrafen1986, Walrafen1996, Zelsmann1995}.
Based on earlier studies on this peak~\cite{Walrafen1996, Nielsen1996, Sharma2005, Chen2008, Sommers2020} and the fact that our study is sensitive only to oxygen atoms, we identify the nature of the HF component as the longitudinal acoustic excitation which, with increasing $Q$, converges to the O--O stretching motion between H-bonded molecules. The decrease of $f$ is then naturally related to the diminishing of the H-bond network.

\begin{figure}
    \includegraphics[width=0.48\textwidth]{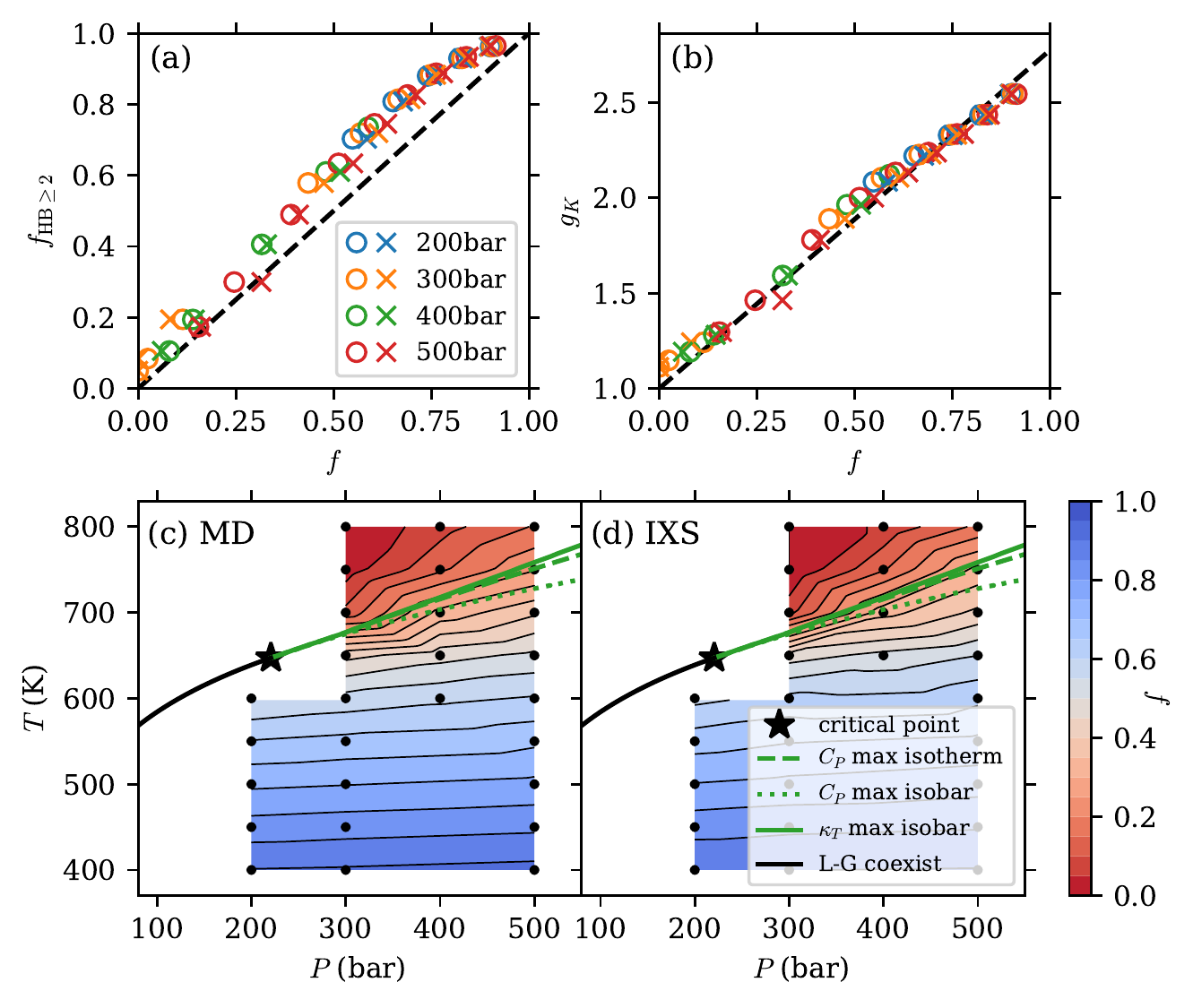}
    \caption{Fraction of the HF component, $f$, plotted against (a) the fraction of molecules with two or more H-bonds, $f_{\text{HB} \geq 2}$,  and (b) the Kirkwood correlation factor $g_K$. Circles: MD; crosses: IXS. Different colors indicate different isobars as shown in the legend. (c) and (d) show contour plots for $f$ on the water phase diagram from MD and IXS, respectively. Black dots indicate states where data is taken; the contour plot is made by linear interpolation on these data. Also plotted are the critical point (black star), liquid-gas coexistence line (black line), and Widom lines by various definitions (green lines; see legend). Thermodynamic data is obtained from the IAPWS-95 equation of state~\cite{Wagner2002}.}
    \label{fig:f_vs_HB}
\end{figure}

In Fig.~\ref{fig:f_vs_HB}(a) we plot the parameter $f$ against the fraction of molecules with two or more H-bonds, $f_{\text{HB} \geq 2}$, obtained from MD results. Here we use a H-bond definition common in the literature: two molecules are H-bonded if their O--O distance is less than \SI{3.5}{\angstrom} and the O$\cdots$O---H angle is less than \ang{30}~\cite{Luzar1996a, Luzar1996}. We note that, even though different definitions result in different H-bond populations~\cite{Matsumoto2007, Strong2018, Kumar2007}, they lead to the same conclusions (see Supplemental Material). 
As shown in Fig.~\ref{fig:f_vs_HB}(a), the data collapse onto a single line with an almost 1:1 ratio, indicating a strong correlation between $f$ and the fraction of $\geq$2-bonded molecules in the probed $P$-$T$ range. We interpret this as evidence to identify molecules giving rise to the HF component with $\geq 2$-bonded molecules which are in the body of the H-bond network, and those contributing to the LF component with monomers and singly bonded molecules in the periphery of the H-bond clusters.

The close relation between $f$ and H-bonding suggests that $f$ may also be connected to chemical properties of sub- to super-critical water. Here we focus on the dielectric constant $\varepsilon$ which controls the solvation behavior and is thus relevant for many chemical processes~\cite{Akiya2002, Weingartner2005}. Notably, $\varepsilon$ decreases from around 80 under ambient conditions to about 6 near the critical point---a value close to that of organic solvents---thus able to dissolve many organic molecules~\cite{Akiya2002, Peterson2008}. A widely used model for $\varepsilon$ is the Kirkwood-Fr{\"o}hlich equation~\cite{Kirkwood1939, Frohlich1949, Suresh2000, Weingartner2005}:
\begin{equation} \label{eqn:K-F_eqn}
    \frac{(\varepsilon - n^2)(2\varepsilon + n^2)}{\varepsilon (n^2 + 2)^2} =  g_K \frac{\rho \mu^2}{9 M \varepsilon_0 k_B T },
\end{equation}
where $\mu = \SI{1.8546}{D}$ is the gas-phase dipole moment of water~\cite{Lide2019} and $\varepsilon_0$ the vacuum permittivity. $n$ is the refractive index at the reference wavelength $\lambda =\SI{0.589}{\micro\metre}$~\cite{IAPWS1997-sp}; we have checked that choosing $\lambda =\SI{1.1}{\micro\metre}$ leads to the same results.
The so-called Kirkwood correlation factor, $g_K$, is defined as the ratio of the total dipole moment in a spherical volume surrounding a fixed molecule to the dipole moment of that molecule, in the absence of external fields. Thus, $g_K=1$ if there is no inter-molecular correlation (as expected for a gas), and $g_K>1$ if molecular dipoles tend to align in parallel (as in room temperature water).
This definition suggests a close relation between $g_K$ and the parameter $f$ in our study. To demonstrate, we calculate $g_K$ using Eq.~\ref{eqn:K-F_eqn} and plot it against $f$ in Fig.~\ref{fig:f_vs_HB}(b). Indeed, good linearity exists between these two parameters. Moreover, the data is consistent with an intercept at $g_K=1$ for $f=0$, where gas-like behavior is expected. Hence we show as a dashed line a linear fit fixing the intercept at $g_K=1$. The fit indicates $g_K = 2.77$ when $f=1$, i.e.\ in the absence of the LF component, which is remarkably close to the value $g_K = 2.79$ at ambient conditions (\SI{1}{bar}, \SI{300}{K}). Therefore, we have established a close relation between $f$ and $g_K$; since the latter is tied to the dielectric constant and is suggested to influence transport properties as well~\cite{Marcus1999}, the parameter $f$ can be viewed as a directly measurable quantity which characterizes chemical properties of supercritical water.

For an overview of the dynamic crossover, in Fig.~\ref{fig:f_vs_HB}(c) and (d) we show contour maps of $f$ on the phase diagram. The topology is clearly influenced by the Widom line which traces the loci of highest gradients, indicating rapid shifts between liquid-like and gas-like dynamics. However, we emphasize that remnants of the HF component can be found above the Widom line and LF below, thus supporting the view of the Widom line as an indicator of continuous crossover instead of a rigid separatrix between distinct liquid-like and gas-like phases~\cite{Schienbein2018}.

In conclusion, we have used a combination of IXS measurements and MD simulations to study the crossover from liquid-like to gas-like dynamics in supercritical water. The TIP4P/2005 model for MD well reproduces the inter-molecular dynamics measured \emph{via} IXS. We find that the dynamics can be separated into liquid-like (HF) and gas-like (LF) components, and it is the changing ratio between the two that leads to the crossover. Through further analysis, we find a strong correlation between the fraction of the HF component and molecules with two or more H-bonds, which is significant for modeling of solvent properties such as the dielectric constant. The Widom line, which originates from thermodynamic properties, coincides with rapid changes in the inter-molecular dynamics and H-bonding as well.

\begin{acknowledgments}
This work is supported by the U.S.\ Department of Energy, Office of Science, Office of Basic Energy Sciences under Contract No.\ DE-AC02-76SF00515. The synchrotron radiation experiments were performed at BL43LXU in SPring-8 with the approval of RIKEN (Proposal No.\ 20180031).
We would like to thank Stanford University and the Stanford Research Computing Center for providing computational resources and support that contributed to this work.
\end{acknowledgments}


%

\end{document}